\documentclass{appolb}
\usepackage{epsfig,amssymb,xspace}
\newcommand{\nc}{\newcommand}
\newcommand{\ave}[1]{\left\langle #1 \right\rangle}
\nc{\req}[1]{Eq.\,(\ref{#1})}  \nc{\eps}{\varepsilon}
\nc{\beq}{\begin{equation}}     \nc{\beql}[1]{\begin{equation}\label{#1}}
\nc{\eeq}{\end{equation}}       \nc{\rf}[1]{figure  \ref{#1}}
\nc{\beqa}{\begin{eqnarray}}   \nc{\eeqa}{\end{eqnarray}}
     \nc{\pathlaptop}{/home/rafelski/figure/}
     \nc{\pathletes}{/users/lpthe/jletes/bookraf/figures/}
     \nc{\pathnow}{}
 \def\lessim{\lower.5ex\hbox{$\; \buildrel < \over \sim \;$}}
\def\gtrsim{\lower.5ex\hbox{$\; \buildrel > \over \sim \;$}}

\usepackage{graphicx}
\begin{document}
\title{Mach Cones in Heavy Ion Collisions%
\thanks{Lecture Notes for the Krakow School of Theoretical Physics, XLVIIIe Course, 2008.
}%
}
\author{
Giorgio Torrieri
\address{Institut f\"ur Theoretische Physik, Goethe-Universit\"at, Frankfurt, Germany}
\and
Barbara Betz
\address{Institut f\"ur Theoretische Physik, Goethe-Universit\"at, Frankfurt, Germany}
\and
Jorge Noronha
\address{Department of Physics, Columbia University, New York, New
York, 10027, USA}
\and
Miklos Gyulassy\address{Department of Physics, Columbia University, New York, New
York, 10027, USA}}
\maketitle
\begin{abstract}
We study the fate of the energy deposited by a jet in a heavy ion collision assuming that the medium created is opaque (jets quickly lose energy) and its viscosity is so low that the energy lost by the jet is quickly thermalized. The expectation is that under these conditions the energy deposited gives rise to a Mach cone. We argue that, in general, the behavior of the system is different from the naive expectation and it depends strongly on the assumptions made about the energy and momentum deposited by the jet into the medium. We compare our phenomenological hydrodynamic calculations performed in a static medium for a variety of energy-momentum sources (including a pQCD-based calculation) with the exact strong coupling limit obtained within the AdS/CFT correspondence.  We also discuss the observability of hydrodynamical features triggered by jets in experimentally measured two-particle correlations at RHIC.
\end{abstract}

\section{Introduction\label{Int}}

One of the most prominent experimental discoveries made at the
Relativistic Heavy Ion Collider (RHIC) has been the suppression of
highly energetic particles \cite{BRAHMS,PHENIX,PHOBOS,STAR}, which
suggests that the matter created at RHIC is a color-opaque, high
density medium of colored degrees of freedom where fast partons quickly lose
energy by gluon emission \cite{Gyulassy:1990ye,Gyulassy:1993hr,Wang:1994fx,Baier:1996kr,Wiedemann:2000za,Gyulassy:1999zd,Wang:2001ifa,Arnold:2001ms,Liu:2006ug,Majumder:2007zh}. Measurements
of anisotropies in soft particle momentum distributions
\cite{BRAHMS,PHENIX,PHOBOS,STAR} have further indicated that soft degrees of freedom are approximately thermalized.  The degree of thermalization has been found to be considerably above the
predictions \cite{gyulvisc} obtained within perturbative Quantum Chromodynamics
(pQCD) and, in fact, it seems to be compatible with
the ``perfect fluid'' scenario where the strongly coupled
Quark-Gluon Plasma (sQGP) has almost zero viscosity
\cite{heinz,shuryak,teaney,heinz2,romatschke,Xu:2007jv,Gyulassy:2004zy,Shuryak:2004cy,Shuryak:2008eq}. These two findings, taken together, suggest that the energy deposited by the jet into the medium is thermalized and becomes part of the fluid.  The theory governing the further evolution of this energy is that of hydrodynamic sound waves.
\begin{figure}[h]
\epsfig{width=12cm,figure=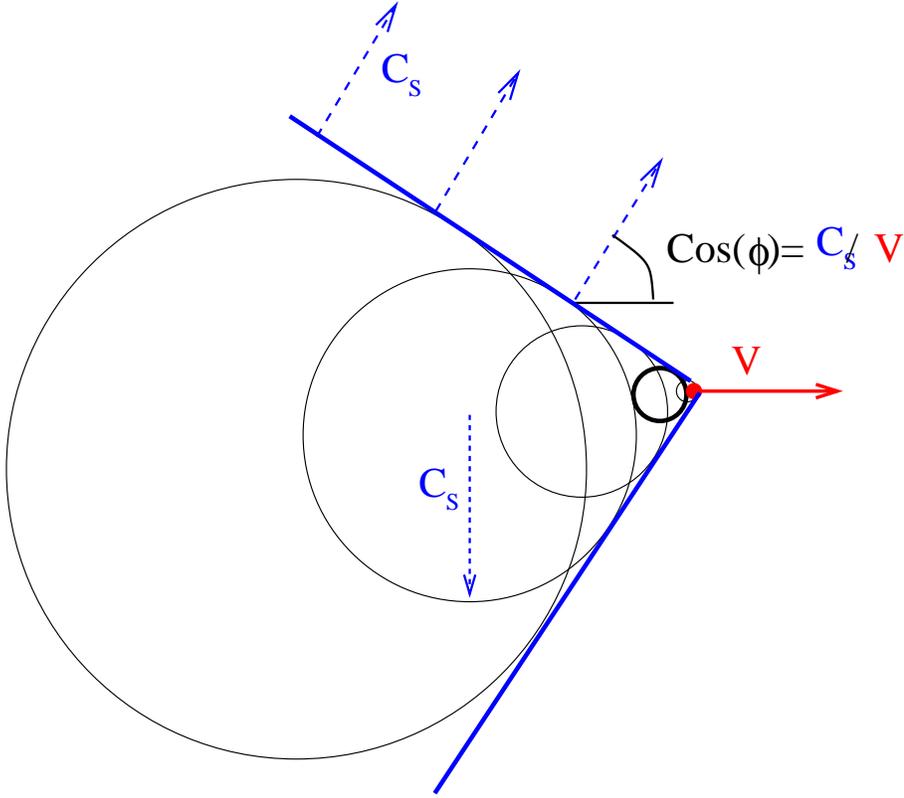}
\caption{(Color online) A geometric sketch of the creation and evolution of a Mach cone.}
\label{machgeom}
\end{figure}
\begin{figure}[h]
\epsfig{file=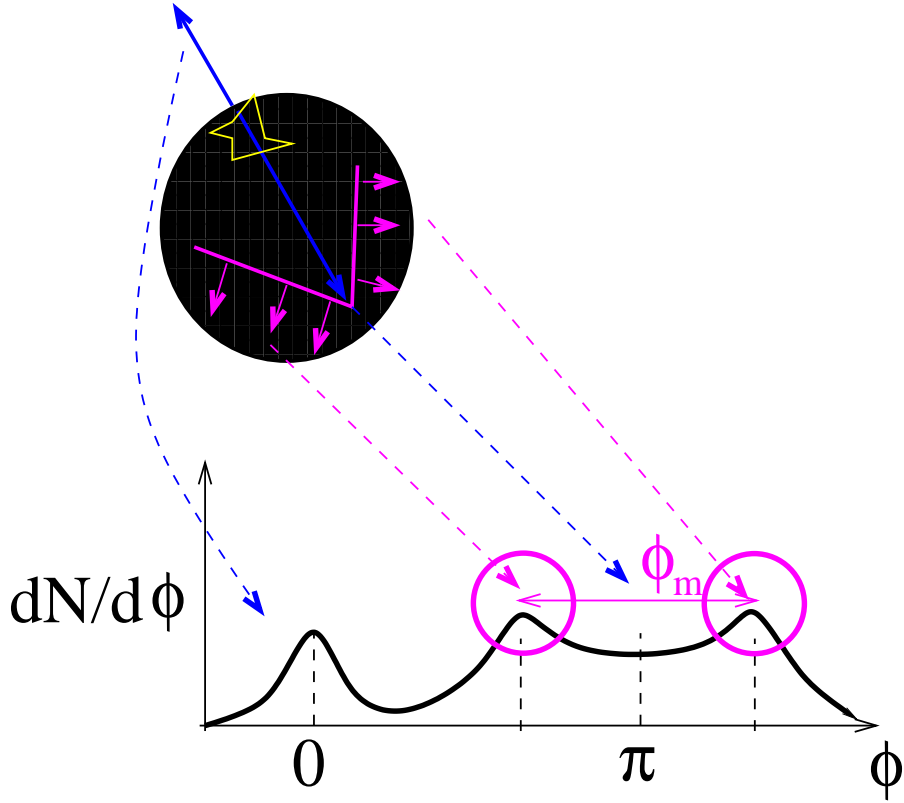,width=12cm}
\caption{(Color online) A schematic representation of a Mach cone in heavy ion collisions.
\label{machscheme}
}
\end{figure}

\begin{figure}[h]
\epsfig{width=12cm,figure=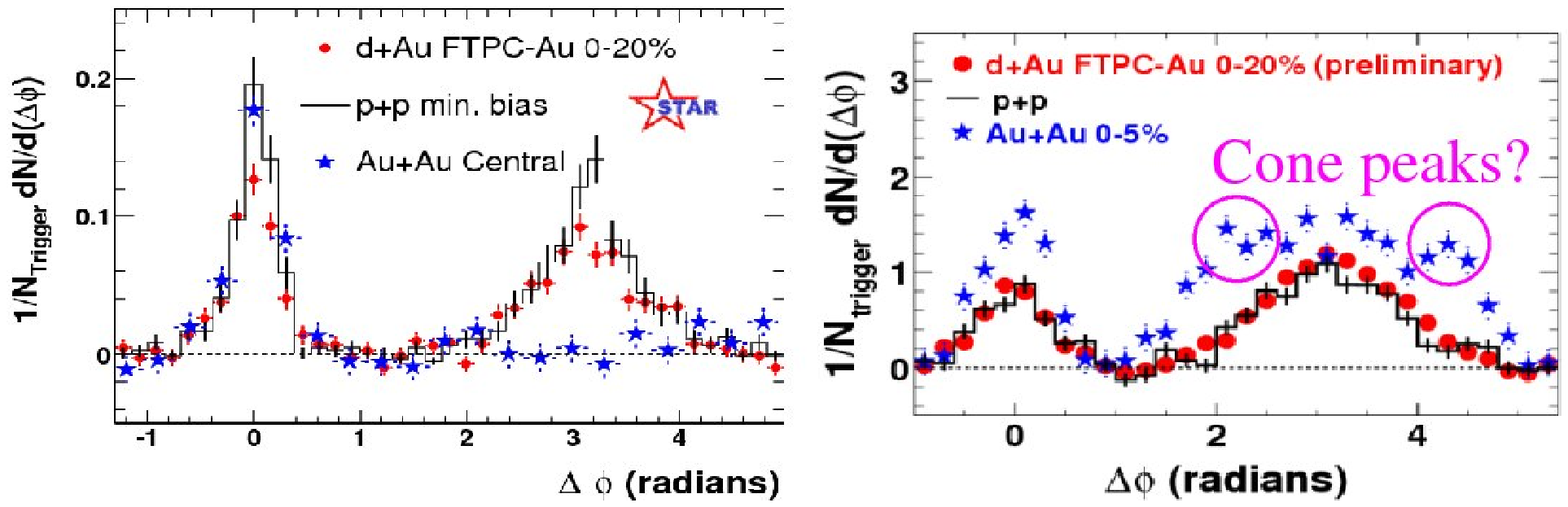}
\caption{(Color online) Two-particle correlations induced by jets in d-Au collisions (red) and Au-Au collisions (blue) at RHIC \cite{expreview,expreview2,expphenix}.  With a hard
away-side trigger (left, $p^{assoc}_T>2$~GeV/c), the away-side peak disappears in agreement with the hypothesis that the jet is absorbed by the medium. However, if the away-side
trigger momentum is lowered (right panel, $0.15>p^{assoc}_T>4$~GeV/c), the peak reappears (as expected from momentum conservation) and shows a cone-like pattern.  The absence of similar correlations in the d-Au ``control'' experiment confirms that this is an effect of the ``medium'' rather than a deviation of the initial conditions.}
\label{jetexp}
\end{figure}

It was noticed in the 19th century that, when a probe travels through a fluid with a speed greater than the speed of sound, the energy deposited by the probe creates a forward moving conical shock-wave.  The reason for this can be seen in Fig.\ \ref{machgeom}: at each point the deposited energy becomes a sound-wave, which moves at the speed of sound $c_s$. The spherical waves interfere coherently creating the shock wave \cite{lifshitzlandau}. Fig.\ \ref{machgeom} can be used, together with a simple geometrical argument, to show that the cone angle is related to the speed of sound $c_s$ and the velocity of the jet $v$ as
\begin{equation}
\label{machangle}
\cos \phi_M = \frac{c_s}{v}
\end{equation}
and, thus, an experimental observation of a Mach cone provides both a proof of the fluid-like behavior of the underlying system and a direct probe of its equation of state. This idea actually predates the current investigations about sQGP and even quark-gluon plasma by decades \cite{stoecker,stoecker2}.

Similarly to ordinary sound waves, Mach cone shocks dissipate exponentially ($\sim e^{- k^2 \Gamma x}$) with respect to the wave-number $k$ and the distance traveled $x$,  where the characteristic sound wave attenuation length $\Gamma$ is related to the shear viscosity $\eta$, the energy density $\varepsilon$, and the pressure $p$
\begin{equation}
\Gamma = \frac{4}{3} \frac{\eta}{\varepsilon+p}.
\end{equation}
Therefore, the presence of a Mach cone signal would be an exponentially precise confirmation of the low viscosity fluid limit, which is in principle more sensitive to $\eta$ than even anisotropic flow.

As with all heavy ion observables sensitive to hydrodynamic behavior, the Mach cone signal suffers from the problem that we do not ``see'' the fluid directly: only the final many-particle correlations are measured and they are sensitive to all stages of the hydrodynamic evolution including the late (presumably non-thermalized) stages and freeze-out (which is not understood from first principles).  A rough approximation is to assume that at a certain locus in space-time $\Sigma^\mu = (t,\vec{x})$ the mean free path goes from zero to infinity.  This locus can be defined in terms of a local criterion (e.g. a common freeze-out temperature), or using a simple global geometry (isochronous freeze-out).   Using Stoke's theorem, as well as entropy, energy and momentum conservation yields the famous Cooper-Frye (CF) formula \cite{cooperfrye}
\begin{equation}
E \frac{dN}{d^3 p} = \int d^3 \Sigma_\mu P^\mu f(U_\mu P^\mu,T)
\end{equation}
where $U^\mu$ is the collective flow vector,  $f(E,T)$ the standard Boltzmann, Bose-Einstein or Fermi-Dirac distribution function (in terms of temperature $T$), and $P^\mu$ is the 4-momentum vector $(E,\vec{p})$ of the associated particle.   Using an azimuthal coordinate system and putting the jet direction at the origin yields a characteristic distribution $dN/d\phi$ with two Mach cone-like peaks shown in Fig. \ref{machscheme}.   Note that ``theoretically'' this is an {\em average} distribution since we {\em define} the near-side jet to be at $\phi=0$.

Experimentally, however, this is a 2-particle correlation: the experiment measures a high momentum ``jet" particle (the near-side trigger) and then looks at the correlation between the trigger and softer (sensitive to flow) particles in the opposite direction (away-side region, where it is assumed that the jet passed through the medium, was suppressed, and its energy thermalized into Mach shocks). Tantalizingly, something similar does seem to be observed in the experiment (Fig.\ \ref{jetexp}):  when a hard (jet-like) particle is correlated with a soft (medium-like) particle, the ``missing'' away-side jet signal reappears and shows a structure very similar to that of a Mach cone with an angle close to the expectation for the ideal gas speed of sound $c_s=1/\sqrt{3}$.

It should be said that this result has generated a lot of skepticism and some debate (for a review see \cite{expreview,expreview2,expphenix}).   The ``dip'' between the Mach cone
peaks arises most clearly when the background 2-particle correlation (arising, for example, from the ellipticity of the initial fireball) is subtracted (ZYAM method).  This subtraction makes possibly unjustified assumptions, e.g., that the Mach cone correlation and the elliptic correlation are independent. Moreover, different theoretical interpretations have been given to the apparent conical structure \cite{noncone1,noncone2,noncone3,noncone4}, which indicates that further studies involving, for instance, 3-particle correlations \cite{Abelev:2008nd} and the dependence of the Mach cone angle on various other variables \cite{Winter:2008bn,Sickles:2008uh} should be pursued.

Nevertheless, the experimental observation of something that looks like a Mach cone has greatly enhanced the theoretical interest on this phenomenon \cite{Stoecker:2004qu,CasalderreySolana:2004qm,shuryakmach,mishustin,heinzmach,chaumach,renk,Friess:2006fk,yarom,Chesler:2007an,gubser2,Gubser:2007ni,
Chesler:2007sv,us1,us2,us3,us4,us5,us6,us7,neufeld,
neufeldsource,neufeld3,Neufeld:2008eg}. In particular, an interesting 
question that still remains to be properly answered is to what extent 
does the ``naive'' Mach cone picture introduced earlier survive when 
realistic physics conditions are introduced.   Recent studies seem to 
indicate the Cone+diffusion wake picture, and the relative weight of the 
two, are remarkably independent of the energy-deposition scenario 
\cite{us8}.

In the next few sections we shall examine this issue in detail. Section \ref{sechydro} describes an analysis of Mach cone propagation using ideal three-dimensional full (non-linear) hydrodynamics, which shows that the observable signal depends crucially on how energy and momentum are deposited by the jet into the medium. Section \ref{secads} uses the Anti-de Sitter/Conformal Field Theory (AdS/CFT) correspondence to try to obtain insights into the non-equilibrium dynamics displayed by this problem, introduces the ``Head + Neck'' decomposition of the space-time region near the jet (based on the degree of thermalization of the medium), and derives the contribution of each region to the particle correlations. Section \ref{secpqcd} does an analogous analysis with an energy-momentum source calculated in pQCD coupled with ideal hydrodynamics and describes the observable differences with respect to results obtained using AdS/CFT.

We use natural units and the Minkowski metric $g_{\mu\nu}={\rm diag}(-,+,+,+)$.
Lorentz indices are denoted with Greek letters $\mu,\nu=0,\ldots,3$. Throughout this paper we shall use a coordinate system corresponding to axial symmetry with respect to the trigger jet axis.   The components of a generic vector $x$ will be
\begin{equation}
x^\mu =\left(\begin{array}{c}t\\ x_1= x-vt\\x_{\perp}\cos \varphi \\ x_{\perp} \sin \varphi \end{array} \right)
\end{equation}
 coordinates, where $x/x_{\perp}$ are taken to be parallel/perpendicular to the near-side jet.   The jet is assumed to be moving transversely with respect to the beam (z axis) so the rapidity of the jet is taken to be zero.  The away-side direction in this system is $(0,1,0,0)$.

\section{Mach Cones in Hydrodynamics \label{sechydro}}

The geometrical argument for Mach cone formation and the angle formula in Eq.\ (\ref{machangle}) are only valid in the {\em linearized} hydrodynamics limit where the energy of the sound waves is small compared to the energy density of the background.    The physical applicability of this condition is doubtful since the order of magnitude of the jet's ``size'' $\sim 1$ fm, which leads to an overwhelming energy density close to it even for moderate rates of energy deposition.   A ``pile-up'' of sound waves in front of the jet as it travels through the medium can also result in strong non-linear corrections to the fluid's flow profile.
\begin{figure}[h]
\epsfig{file=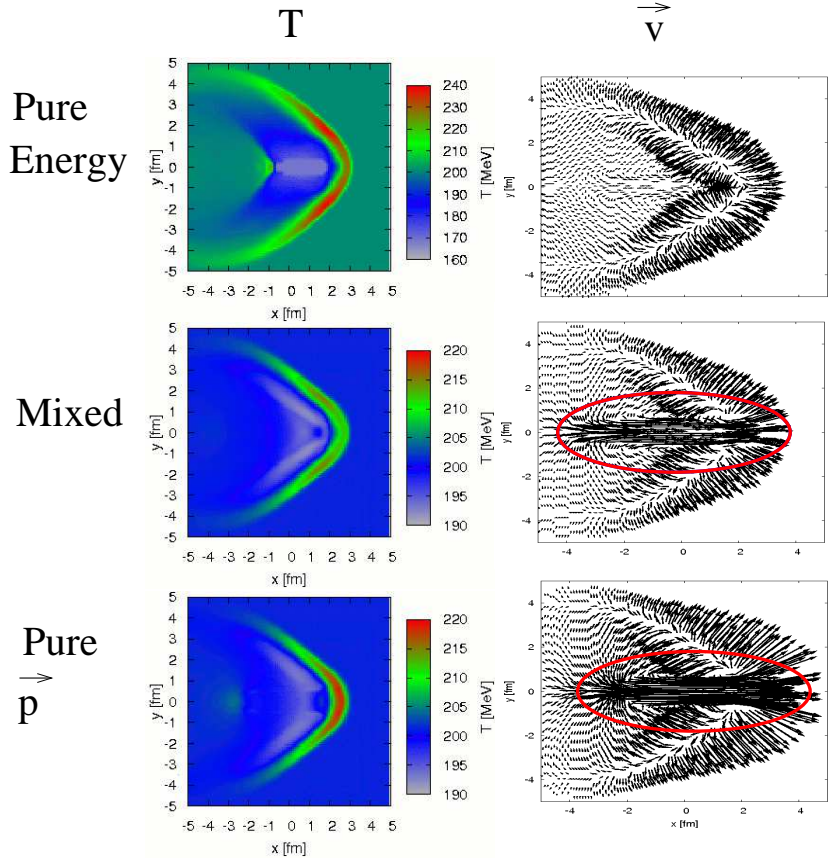,width=12cm}
\caption{(Color online) Temperature (left panel) and flow (right panel) profiles for a hydrodynamic simulation of a jet with different energy and momentum deposition schemes in a static medium (from Ref.\ \cite{us2}).}
\label{diffwake}
\end{figure}
Furthermore, if the energy deposition is {\em small} with respect to the background, an observable signal will not survive the {\em thermal} background fluctuations inherent in a
CF freeze-out. It is relatively easy to show {\em analytically} \cite{shuryakmach,us5} that this is the case unless {\em either} the strength of the signal is large enough to require a non-linearized treatment {\em or} the momentum is large enough to put the thermalization assumption in doubt \footnote{``Harder'' particles are more contaminated by non-thermal processes such as minijet fragmentation.  While this is ``obvious'' since the effective mean free path grows with the particle momentum, the scale at which the medium stops being the main source of particles is not known precisely (though assumed to be in the $p_T\sim 1-2$ GeV range).}.

For associated (massless) particles with
$P^{\mu}=(p_{T},p_{T}\cos (\pi-\phi),p_{T}\sin
(\pi-\phi),0)$
the momentum distribution at mid rapidity is given by
\begin{equation}
\frac{dN}{p_Tdp_T d\phi}\Big
|_{y=0}=\int_{\Sigma}d\Sigma_{\mu}P^{\mu}\left[f(U^{\mu},P^{\mu},T)-f_{eq} \left( P^0,T_0  \right)\right]
\label{coopermach}
\end{equation}
We subtract  a non-flowing $U^{\mu}=(1,0)$ thermal $T=T_0$ constant background yield via $f_{eq}\equiv f|_{U^{\mu}=0,T=T_0}$. Viscous corrections to the Boltzmann distribution function \cite{Dusling:2007gi} produce subleading contributions that are negligible in the linearized approximation.  Choosing an isochronous ansatz where $d\Sigma^\mu= x_\perp d x_\perp dx_1 d\varphi\,\left(1,0,0,0\right)$, the Boltzmann exponent can be expanded
up to corrections $\mathcal{O} (\ave{U}^4)$. The associated away side azimuthal distribution at mid-rapidity $f(p_T,\phi)=dN/p_{T}dp_{T}dyd\phi |_{y=0}$ with respect to the nuclear beam axis is then given after integrating over $\varphi$ by
\begin{eqnarray}
f(p_T,\phi)=2\pi\,p_T\,\int_{\Sigma} dx_1 dx_\perp x_\perp \times
\label{fulldistrib}
\end{eqnarray}
\[\ \left(
\exp\left\{-\frac{p_T}{T}\left[U_0-U_1\cos(\pi-\phi)\right]\right\}\, I_0(a_\perp)-e^{-p_T/T_0} \right) \]
where $a_\perp=p_T U_\perp\sin(\pi-\phi)/T$ and $I_0$ is the modified Bessel function. In the linearized approximation $a_{\perp} \ll 1$ and, thus, we can use the expansion for the Bessel function
\begin{equation}
\label{bessel}
\lim_{x\to 0}\,I_0 (x) =1+\frac{x^2}{4}+\mathcal{O}(x^4)
\end{equation}
to get the approximate equation for the distribution
\begin{equation}
f(p_T,\phi) \simeq e^{-p_{T}/T_0}\frac{2\pi\,p_{T}^2}{T_0}
\left[\frac{\langle \Delta T\rangle}{T_0} + \langle U_{1}\rangle \cos
(\pi-\phi)  
\right]\,.\label{expandedCFfinal}
\end{equation}
Note that deviations from isotropy are then controlled by the following global moments $\langle \Delta T\rangle= \int_{\Sigma} dx_1 dx_\perp x_\perp \,\Delta T$ and $\langle U_1\rangle= \int_{\Sigma} dx_1 dx_\perp x_\perp \,U_1$.

It is clear that in the strict linearized limit the azimuthal distribution only has a trivial broad peak at $\phi=\pi$. A double-peaked structure in the away-side of the jet correlation function can only arise when the Bessel function expansion is invalid, i.e., away from the linearized limit or for large $p_T \gg T_0$. For such large momenta, contamination from non-thermalized degrees of freedom and coalescence \cite{molnar,molnarcoal,Fries:2003vb,Fries:2003kq,Fries:2004hd,Greco:2003xt,Greco:2003mm} effects are non-negligible.

On the other hand, for a non-linear perturbation in the medium it is not clear that the Mach cone angle is anything like the one derived geometrically in the Introduction. For large energy depositions, the more appropriate description is that of an angular {\em shock}, i.e., a step function in energy density.  This problem was analyzed analytically in \cite{dirkmach}, where the angle was found to be in general larger than the geometrical expectation from Mach's law. Deviations from equilibration, expected in the region where the energy deposition from the jet is comparable to the local energy density, may also lead to a different cone angle.

In addition, the Mach cone is not the {\em only} collective mode that 
can be excited by the moving jet. The ratio of momentum to energy 
deposition is not known but a common assumption is to have the same 
amount of energy and momentum being deposited by the jet. Longitudinal 
momentum deposition results in a column of fluid that flows in the 
direction of the jet's motion. This structure is known as a the 
``diffusion wake''. We have taken all these effects into account in 
Ref.\ \cite{us2} where we studied energy and momentum deposition in a 
static medium using the 3D ideal hydrodynamics SHASTA code presented in 
\cite{dirk2}. Note that the simulations presented below only refer to a 
static thermal background, and as such neglect the effects of transverse 
and longitudinal flow (which are a simple deformation in the linearized 
approximation \cite{mishustin} but can be non-trivial in full nonlinear 
hydrodynamics \cite{us8}).

\begin{figure}[h]
\begin{center}
\epsfig{file=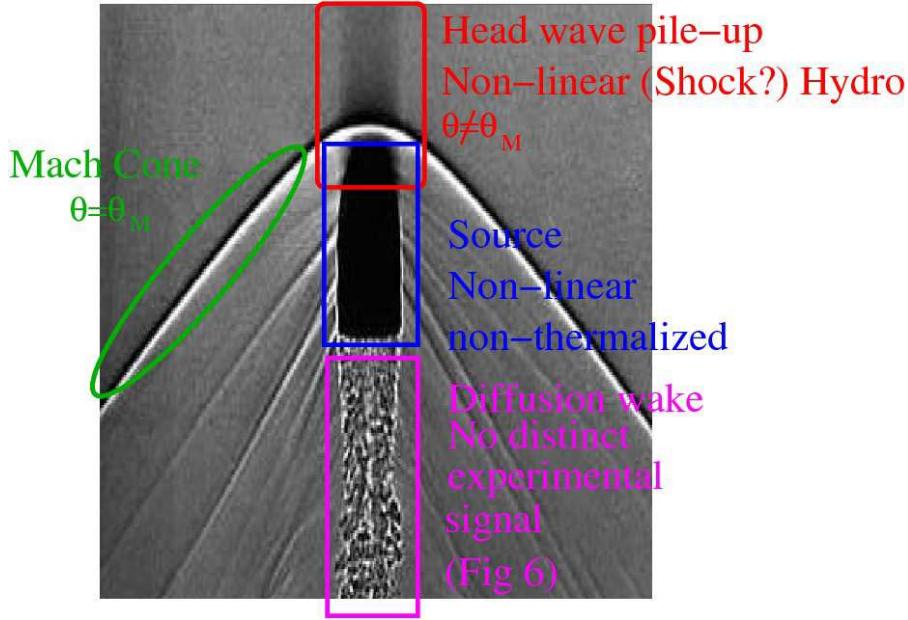,width=12cm}
\caption{(Color online) A schematic representation of a Mach cone solution including non-linear and non-equilibrium effects.
\label{machdetail}
}
\end{center}
\end{figure}

Fig. \ref{diffwake} shows what happens when the momentum deposition is 
included in the full hydrodynamic simulations \cite{us2}:  the temperature profile remains invariant and maintains the correct angle.  However, the flow profile acquires an additional component co-moving with the jet. See Fig. \ref{machdetail} for a sketch of all these effects combined.  It then becomes clear that the appearance of a cone-like signal {\em in hydrodynamics} is {\em not} assured, since the strongest signals may come from contributions which are not conical and possibly not locally equilibrated. More complicated momentum depositions, such as including transverse momentum, or introducing a momentum dependence of the deposited energy (the so-called ``Bragg peak'') do not change these conclusions 
qualitatively \cite{us8}.

\begin{figure}[h]
\epsfig{file=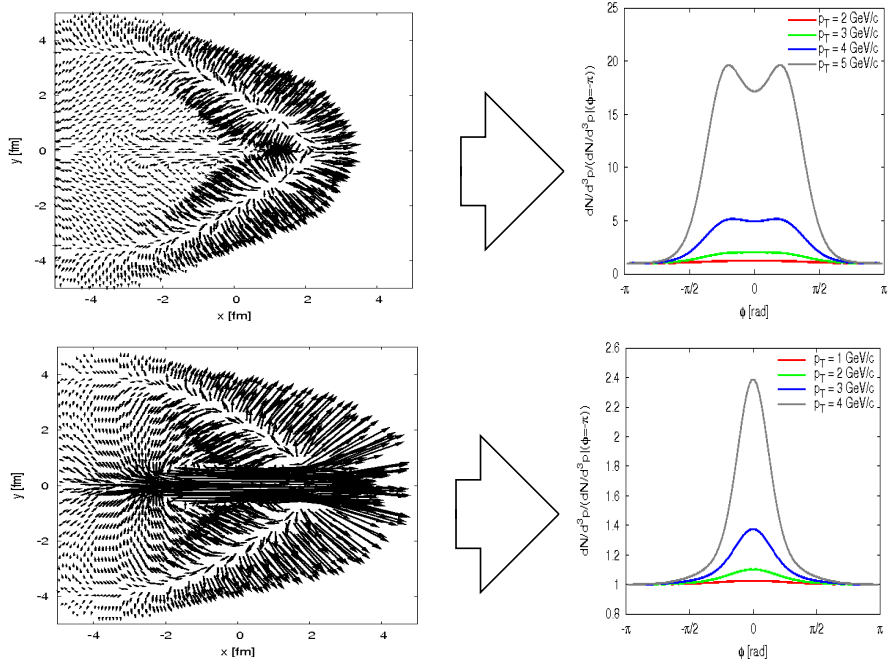,width=12cm}
\caption{(Color online) \label{barbarafo} Effect of freeze-out for various energy-momentum depositions \cite{us2}. Note that angle $\phi$ is shifted in the left-hand plot in such a way that the away-side peak is located at $\phi=0$.
}
\end{figure}

-Fig. \ref{barbarafo} shows what happens after freeze out: when conical distributions dominate, the system does exhibit a cone signal with the correct angle, albeit the momenta of the associated particles are so high that one is likely to question their effective thermalization once the ideal fluid approximation is relaxed. On the other hand, when the diffusion wake is significant the only correlation visible is a unique peak in the away-side, which is indistinguishable from a generic peak expected from momentum conservation.

To strengthen this conclusion, note that the pure energy deposition scheme may be completely undetectable in a realistic and approximately isothermal freeze-out condition since hotter regions will simply freeze-out {\em later}. Momentum flow, however, will persist independently of freeze-out and will in general be modified only a little by the last (cooler) stages of the fireball evolution.  Thus, it is the right panel of Fig.\ \ref{diffwake} that shows the fluid correlations most apt to be imprinted on two-particle correlation functions.

In sum, a phenomenological hydrodynamical approach shows that Mach cones are not {\em guaranteed} to appear in the final angular correlation functions. The problem of the away-side jet correlations in the sQGP is a very interesting and complicated subject because of its inherent out of equilibrium, non-linear, and non-perturbative features. An interesting development in this direction was provided by the possibility of using string theory dualities to calculate observables in a strongly coupled gauge theory without the ``a priori" equilibration assumptions.

\section{Mach Cones in AdS/CFT \label{secads}}

\subsection{Introduction}

The energy-momentum tensor of a system composed of a heavy quark passing through a strongly coupled $\lambda\gg 1$ ($\lambda=g_{SYM}^2 N_c$ is the t'Hooft coupling) $\mathcal{N}=4$ $SU(N_c)$ Yang-Mills plasma at finite temperature $T$ can be computed using the AdS/CFT correspondence \cite{Friess:2006fk,yarom,Chesler:2007an,gubser2,Gubser:2007ni,
Chesler:2007sv}, first conjectured in \cite{maldacena}. According to the correspondence, gauge invariant observable quantities in a strongly coupled $\mathcal{N}=4$ SYM theory can
be determined using weakly coupled 10 dimensional type IIB superstring theory, where 5 of the dimensions are Anti-de Sitter and the other 5 correspond to a 5-sphere.

\begin{figure}[t]
\epsfig{file=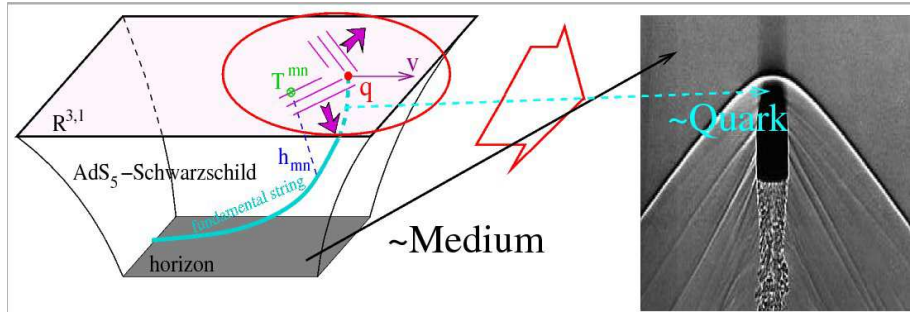,width=12cm}
\caption{(Color online) The Mach cone set-up associated with heavy quarks within the AdS/CFT correspondence (the figure on the left-hand side was taken from Ref.\ \cite{Friess:2006fk}).  The finite-temperature medium corresponds to an event horizon and the quark to the tip of a fundamental string stretched between the horizon and the boundary of $AdS_5$.
\label{machads}
}
\end{figure}

The problem of a heavy quark \footnote{Note that ``Mach-like'' signals found in experiment so far are triggered by light quark/gluon jets  \cite{PHENIX,STAR,expreview,expreview2}.} moving at constant speed in a strongly coupled finite temperature  $\mathcal{N}=4$ SYM medium can be analyzed (Fig.\ \ref{machads}) by considering metric fluctuations due to a string that
is hanging down from the boundary of an AdS Schwarzschild (AdS-SS) background
geometry \cite{Herzog:2006gh,gubser4,CasalderreySolana:2006rq}. Quantum fluctuations can be neglected for a slowly moving heavy quark \cite{Herzog:2006gh}.  In this limit, the action that describes the supergravity approximation to type IIB string theory in an AdS-SS background and the classical string is given by the sum of the following partial actions
\begin{equation}
\ A_G = \frac{1}{16\pi G_5 }\int \sqrt{-G}\left(R+\frac{12}{L^2} \right)
\label{gravityaction}
\end{equation}
and
\begin{equation}
\ A_{NG} = -\frac{1}{2\pi \alpha' }\int \sqrt{-G_{\mu\nu}^{(0)}\partial_{\alpha}X^{\mu}\partial_{\beta}X^{\nu}}d^2\sigma,
\label{nambugoto}
\end{equation}
where $L$ is the radius of $AdS_5$, $G_5=\pi L^2 /2N_{c}^2$, $\alpha'=L^2 /\sqrt{\lambda}$, $G_{\mu\nu}$ is the total metric, and $G_{\mu\nu}^{(0)}$ is the metric of the unperturbed AdS-SS black hole (without effects from the string), which can be obtained from
\begin{equation}
\ ds^2 = \frac{L^2}{z^2}\left(-g(z)dt^2+d\vec{x}^{\,2}+\frac{dz^2}{g(z)}\right)
\label{lineelement}
\end{equation}
where $z$ goes from $0$ at the AdS boundary to $z_0 =1/\pi T$ at the black hole horizon ($T$ is the Hawking temperature associated with the black hole) and $g(z)=1-(z/z_0 )^4$. The string coordinates $X^{\mu}(\sigma,\tau)$ in the Nambu-Goto action in Eq.\ (\ref{nambugoto}) are chosen in such a way that the string endpoint (which corresponds to the heavy quark in the 4-dimensional boundary) moves at constant speed $v$ and no energy flows from the horizon into the string \cite{Herzog:2006gh,gubser4}.

Minimizing the action $S$ with respect to the metric $G$ leads to the full set of Einstein's equations. It is sufficient for our purposes here to consider instead the linearized Einstein's equations for the metric fluctuations $h_{\mu\nu}$, which are defined via $G_{\mu\nu}=G_{\mu\nu}^{(0)}+h_{\mu\nu}$. It can be shown \cite{Friess:2006fk,yarom} that the contribution from the moving quark to the total energy-momentum tensor\footnote{By $T_{\mu \nu}$ we mean $\ave{T_{\mu \nu}}$, though we will drop the $\ave{...}$ notation for brevity.} is $T_{quark}=\frac{1}{\pi}\sqrt{\frac{\lambda}{1-v^2}}Q$, where the tensor $Q$ is obtained by expanding $h$ in powers of $z$ near the boundary, i.e., $h\sim Q\,z^4$. An example of the formidable analytical power of AdS/CFT calculations was given by Yarom in Ref.\ \cite{yarom} (see also \cite{Gubser:2007nd}) where he computed the total energy-momentum tensor in the lab frame that describes the near-quark region
\begin{equation}
\ T_{\mu\nu}^{Y}=P_0\,{\rm diag}\{3,1,1,r^2\}+\xi\,P_0 \,\Delta T_{\mu\nu}(x_{1},r)
\label{energymomentumtensor1}
\end{equation}
where the explicit form of $\Delta T_{\mu \nu}$ is
\begin{eqnarray}
 \Delta T_{tt} & = \alpha \frac{v  \left(r^2(-5+13v^2-8v^4)+(-5+11v^2)x_1^2\right)x_1}{72\left(r^2(1-v^2)+x_1^2\right)^{5/2}}, \label{yarmom1}\\
 \Delta T_{t x_1} & = -\alpha \frac{v^2 \left(2 x_1^2+(1-v^2)r^2\right)x_1}{24 \left(r^2(1-v^2)+x_1^2\right)^{5/2}}, \label{yarmom2}\\	
 \Delta T_{tr} & = -\alpha \frac{(1-v^2) v^2 \left(11 x_1^2+8r^2(1-v^2)\right) r}{72 \left(r^2(1-v^2)+x_1^2\right)^{5/2}}, \label{yarmom3}\\	
 \Delta T_{x_1 x_1} & = \alpha \frac{v  \left(r^2(8-13 v^2 +5v^4)+(11-5v^2)x_1^2\right)x_1}{72 \left(r^2(1-v^2)+x_1^2\right)^{5/2}}, \label{yarmom4}\\
 \Delta T_{x_1 r} & = \alpha \frac{ v (1-v^2) \left(8 r^2(1-v^2)+11 x_1^2\right)r}{72 \left(r^2(1-v^2)+x_1^2\right)^{5/2}}, \label{yarmom5}\\
 \Delta T_{r r} & = -\alpha \frac{v (1-v^2)  \left(5 r^2(1-v^2) + 8 x_1^2\right)x_1}{72 \left(r^2(1-v^2)+x_1^2\right)^{5/2}}, \label{yarmom6}\\
 \Delta T_{\theta \theta} & = -\alpha \frac{v (1-v^2) x_1 }{9 \left(r^2(1-v^2)+x_1^2\right)^{3/2}}. \label{yarmom7}
\end{eqnarray}
and $\alpha=\gamma\sqrt{\lambda} T^2$. In the equations above we used dimensionless coordinates in which the distance is normalized to $1/(\pi T)$  and also defined $\xi=8\sqrt{\lambda}\,\gamma/N_{c}^2$ and $\gamma$ as the quark time dilation factor $1/\sqrt{1-v^2}$. Moreover, $P_0=N_{c}^2 \pi^2 T^4/8+\mathcal{O}(N_{c}^0)$ is the pressure of the ideal SYM plasma \cite{Gubser:1996de}.

Note that it is assumed throughout the derivation of Eq.\ (\ref{energymomentumtensor1}) that the metric disturbances caused by the moving string are small in comparison to the $AdS_{5}$ background metric. Therefore, this result is correct as long as this condition is fulfilled. In fact, since $\Delta T_{\mu \nu}$ scales inversely with the total distance from the quark, the region where the condition $\xi\,\Delta T_{\mu \nu} < 1$ (or, equivalently, $h$ small in comparison to $G^{(0)}$) holds can be taken to be arbitrarily small as long as the limit where $N_c\to \infty$ and $\lambda$ is large is employed. However, in order to evaluate the relevance of this approach to heavy ion collisions, one could set $N_{c}=3$, $\lambda=3\pi$ ($\alpha_s =0.25$), and $\gamma =10$, which gives $\xi > 30$. This value of $\xi$ then sets a lower bound of $5/\pi T$ on the minimum distance from the quark that marginally fulfills the condition $\xi\,\Delta T_{\mu \nu} < 1$.

We stress that the energy-momentum tensor shown above is {\em not} a solution of the hydrodynamic equations but rather the {\em full non-equilibrium result} in the strong coupling limit. The resemblance of this set-up to the hydrodynamic calculations sections \ref{Int} and \ref{sechydro} indicates that, for strongly coupled field theories, jet energy deposition is really reduced to something that looks hydrodynamical and linear reasonably close $\sim 5/(\pi T)$ from the jet.

\subsection{Comparison to Hydrodynamics}

The disturbances in the fluid caused by the moving jet are expected to behave hydrodynamically in the region sufficiently far from where the jet is presently located. However, in the near zone close to the heavy quark hydrodynamics must break down and this can be checked explicitly by looking at the isotropy in the Landau frame  (denoted henceforward by brackets and subscript $L$, $\left(...\right)_L$) at each point \cite{us5}, which is defined by the condition that  $(T_{0 i}^{Y})_{L} = 0$.   The boost with respect to the lab frame defines the hydrodynamic flow vector  $U^{\mu}$. Note that, unless the system is a {\em coherent} field where the phase velocity is equal to the speed of light (such would be the case of an electromagnetic wave), this transformation is always possible. This can be accomplished by solving a system of two equations for the two space-like components of $U^\mu$, $U_{1}$ and $U_{r}$ ($U_{\theta}=0$),
\begin{equation}
\left( T_{0 i}^{Y} \right)_L = \Lambda^{\mu}_{i}\,T_{\mu \nu}^{Y}\,\Lambda^{\nu}_{0} = 0 ,
\label{landauequation}
\end{equation}
where $\Lambda^{\mu}_{i}$ is a general coordinate dependent Lorentz transformation
\begin{equation}
\Lambda = \left( \begin{array}{cc} \gamma &  \vec{U}^{T} \\
\vec{U} & \,\,\,\,1+\frac{\vec{U}\otimes\vec{U}^{T}}{\vec{U}^{2}}\left(
\gamma -1 \right)
\end{array}
 \right),
\end{equation}
and $\gamma\equiv U^0=\sqrt{1+\vec{U}^2}$. Using the representation
\begin{equation}
\ T^{Y} = \left( \begin{array}{cc} \varepsilon &  -\vec{S}^{T} \\
-\vec{S} & \,\,\,\,\hat{\tau}
\end{array}
 \right)
\end{equation}
where $\varepsilon=T_{00}^{Y}$ and $\hat{\tau}_{ij}=T_{ij}^{Y}$ ($i,j=1,2,3$), we obtain that Eq. (\ref{landauequation}) becomes
\begin{equation}
\vec{U}=\frac{1}{\left(\gamma\varepsilon-\vec{S}^T \cdot \vec{U}\right)}\left[1+  \frac{\vec{U}\otimes\vec{U}^{T}}{\vec{U}^{2}}\left(
\gamma -1 \right)  \right] \left(\gamma \vec{S}-\hat{\tau}\vec{U}\right).
\label{numericalsolutionflow}
\end{equation}
For finite $N_c$ and $\lambda$ this equation can only be solved numerically. However, for very large $N_c$ and large $\lambda$ one can use that, to leading order in $1/N_c$, $\gamma\approx 1$, $\varepsilon\approx 3P_0$, and $\hat{\tau}\approx P_0$. Using these approximations, one can then obtain that
\begin{equation}
\vec{U}\approx\frac{\vec{S}}{4P_0}.
\label{largeNcflow}
\end{equation}
In deriving the equation above we used that $\xi\sim \mathcal{O}(\sqrt{\lambda}/N_{c}^2)$. The fact that $|\vec{U}| \sim \mathcal{O}(\sqrt{\lambda}/N_{c}^2)$ in the large $N_c$, $\lambda$ limit implies that nonlinear terms in the hydrodynamic description are subleading contributions that can be neglected. Thus, if the system's dynamics can be described by hydrodynamics, these equations have to be linear for the present string theory setup obtained in the large $N_c$, $\lambda$ limit. However, at finite $N_c$ and $\lambda$ nonlinear effects are expected to be relevant. These nonlinear effects can only be properly taken into account by incorporating subleading $1/N_c$ corrections.

 \begin{figure}[h]
 \begin{center}
  \epsfig{width=8cm,clip=,figure=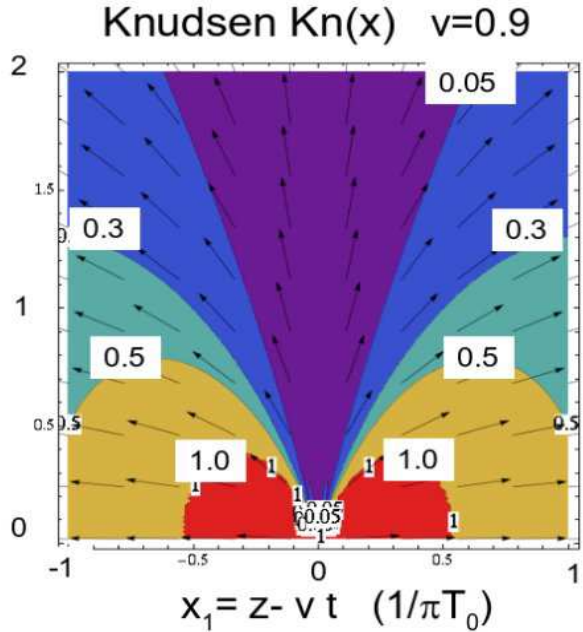} \caption{\label{figknudsen}
 (Color online) The local Knudsen number field for the near zone Yarom stress. Note the $3Kn > 1$
region that defines the Knudsen Neck zone when $v=0.9$.}
 \end{center}
 \end{figure}

In hydrodynamics, the Knudsen number $K_n$ is defined as the ratio between the mean free path $l_{MFP}$ and a characteristic spatial dimension of the system $q$. Hydrodynamics is applicable when $K_n \equiv l_{MFP}/q \ll 1$. In conformal field theories at finite temperature, the only dimensionful parameter is given by the temperature $T$ and, thus, both $l_{MFP}$ and $q$ should be proportional to $1/T$. However, the mean free path is not a well defined quantity in $\mathcal{N}=4$ SYM theories at very strong coupling. Nevertheless, one can still define an effective Knudsen field in terms of the sound attenuation length $\Gamma$ and the Poynting vector $\vec{S}$ as follows \cite{us1}
\begin{equation}
Kn = \Gamma \left| \frac{\nabla.\vec{S}}{S} \right|.
\end{equation}
It is easy to check that in the weak coupling limit this definition reduces to the usual one corresponding to the mean free path over the macroscopic system size (in this case the scale is defined by the flow gradient) or, in other words, the number of times a microscopic degree of freedom interacts while traversing a macroscopic region the system. Also note that this definition is well defined in the supergravity limit. As Fig. \ref{figknudsen} shows, when $v=0.9$ the region where hydrodynamics provides a good approximation of the near-quark region approximately coincides with the locus defined by the local Knudsen number being $Kn^{-1}\geq 3$, which then corresponds to the onset of hydrodynamics ($Kn<<1$) as a sensible approximation. We will later use this condition to further analyze our AdS/CFT results and compare it to a pQCD-like solution. A direct comparison of the near-zone Yarom tensor and a first-order Navier-Stokes ansatz showed \cite{us1} that a hydrodynamic description of the disturbances caused by the heavy quark is valid down to distances of about $1/T$ from the heavy quark.

\subsection{Observability of the AdS/CFT Mach Cone \label{secfreezeads}}

The fact that in the supergravity approximation $T^{\mu\nu}$ can be described by (mostly) linearized hydrodynamics means that once the system breaks up into particles, a conical signal in the corresponding angular correlations may be washed out by thermal smearing, as discussed in the beginning of Section \ref{sechydro}. This means that a detectable Mach cone-like signal may also come from the region where the linearized approximation is {\em not} valid.  In fact, we will show that the only detectable cone-like signal from the AdS/CFT solution comes precisely from the region that is not fully thermalized.

\begin{figure}[h]
\epsfig{width=12cm,figure=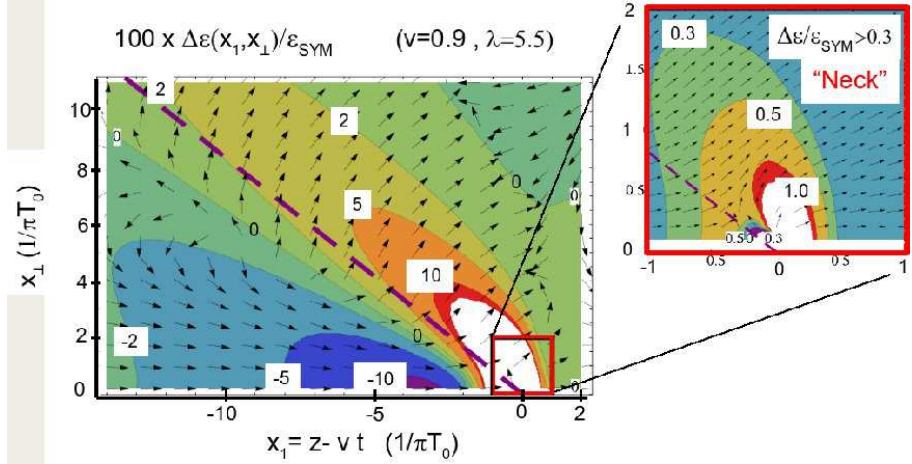}
\caption{\label{adscftmach}(Color online) $T_{00}$ (contour) and $T_{0i}$ (arrows) for the Mach cone AdS/CFT solution (the full numerical computation of \cite{gubser2}).  The three regions defined in Eq. (\ref{headneckmach}) are identified. Note the presence of a strong non-hydrodynamic transverse energy flow near the core. The dashed line shows the Mach cone line in the linearized approximation.}
\end{figure}

The $T^{00}$ and $T^{0i}$ components of the energy momentum tensor (computed within the supergravity approximation) that describe both the near region and the far zone were computed numerically by Gubser, Pufu, and Yarom in Ref.\ \cite{gubser2}. In Fig.\ \ref{adscftmach} we show the energy density perturbation $\Delta\varepsilon(x_1, x_\perp)/\varepsilon_{SYM}$ computed using the data from Ref.\ {\protect\cite{gubser2}} due to a heavy quark jet with $v=0.9$ in a $\mathcal{N}=4$ SYM plasma modeled via the AdS/CFT string drag model for $N_c=3,\;\lambda=g^2_{YN}N_c=5.5$. The left panel shows the far zone (the numbers in the plot label the contours, in per cent as defined on the upper-left corner). The Mach wake zone is above the dashed line, $\cos\phi_M=1/(\surd 3 v)$, and the Diffusion zone lies below that line. Normalized Poynting (momentum flux) vector flow directions are indicated by arrows. The insert shows the nonequilibrium ``Neck'' zone (with the Coulomb Head subtracted) as defined by the condition that $\Delta\varepsilon/\varepsilon_{SYM}>0.3$.

In order to understand the various regions mentioned above, we decompose $T_{\mu \nu}$ into terms that dominate in different spatial
scales (see Fig. \ref{adscftmach}):
\begin{equation}
\label{headneckmach}
T^{\mu\nu}(x)=T^{\mu\nu}_0+\delta T^{\mu\nu}_{Mach}
+ \delta T^{\mu\nu}_{Neck}+ \delta T^{\mu\nu}_{Coul}
\end{equation}
where the far zone ``Mach'' part of the stress tensor is defined by $Kn^{-1}>3$, and coincides with the hydrodynamic description,
\begin{eqnarray}
\label{hydrotensor}
\delta T_{Mach}(x_1,x_\perp) &=& \frac{3}{4}
K\left\{T^4\left(\frac{4}{3}U^\mu U^\nu-\frac{1}{3}g^{\mu\nu}
+ \frac{\eta}{sT} \partial^{\{ \mu}U^{\nu\}} \right)- T^{\mu\nu}_0\right\} \nonumber \\ &\times&
\theta(1-3Kn)
\end{eqnarray}
and the Neck zone is defined by the region close to the heavy quark jet where the local Knudsen field is large and  even uncertainty bounded equilibration rates are too small to maintain local equilibrium. The background stress tensor in $T_0^{\mu\nu}$. As shown in \cite{yarom,us5} the non-equilibrium zone is characterized by a stress of the form
\begin{equation}
\delta T_{Neck}(x_1,x_\perp)\approx \theta(3Kn(x)-1)
\frac{\surd \lambda T_0^2}{x_\perp^2 +\gamma^2 x_1^2}Y^{\mu\nu}(x_1,x_\perp)
\end{equation}
where $Y^{\mu\nu}$ is a dimensionless ``angular'' tensor
field. At very small distances from the jet,
$\delta T_{Neck}(x_1,x_\perp)$ reduces to the analytic stress tensor defined in Eqs. (\ref{yarmom1}-\ref{yarmom7}).

Within the Neck zone, there is also an inner ``Head'' region
where the stress becomes dominated by the contracted
Coulomb self field stress of the quark $\delta T^{\mu\nu}_{Coul}$.
The Head zone can formally  be defined as in
Ref. \cite{Dominguez:2008vd} by equating the analytic
Coulomb energy density \cite{Friess:2006fk}, $\varepsilon_C(x_{1},x_{\perp})$,
to the analytic near zone energy density \cite{yarom} given by Eq. (\ref{yarmom1}).   This Coulomb head boundary is approximately given by
\begin{equation}
x_{\perp}^2 + \gamma^2 x_{1}^2 =
\frac{1}{(\pi T_0)^4}\frac{(2x_{\perp}^2+x_{1}^2)^2}{\gamma^4 x_{1}^2(x_{\perp}^2/2 + \gamma^2 x_{1}^2)^2}.
\end{equation}

The Head zone is  a Lorentz contracted
pancake with longitudinal thickness $\Delta x_{1,C} \,\pi\, T_0 \sim 1/\gamma^{3/2}$
and an effective transverse radius
 $ \Delta x_{\perp,C}\,\pi\, T_0 \sim 1/\gamma^{1/2}$, which is in agreement with the general considerations in
Ref. \cite{Dominguez:2008vd}.  However, as we have shown in Fig. \ref{figknudsen}, for relativistic jets the Neck has a two lobe structure.  The lobes are  nearly independent of $v$ and
the lobe region thickness is
$\Delta x_{1,Kn} \sim 1/\pi T_0\gg \Delta x_{1,C}$.
The second thin pancake component of the $Kn$ that develops for large $\gamma$
(not shown) is similar to the shape of the Head zone.
The relative independence of the two lobe component
of the Neck zone on $v$ is
in agreement with the parametric dependence $\Delta x_{1,N}\propto 2/\pi T_0 \sim 6\Gamma$ expected from the bound of dissipation rates imposed by the uncertainty principle.

\begin{figure}[b]
\begin{center}
\centerline
\centering
\epsfig{file=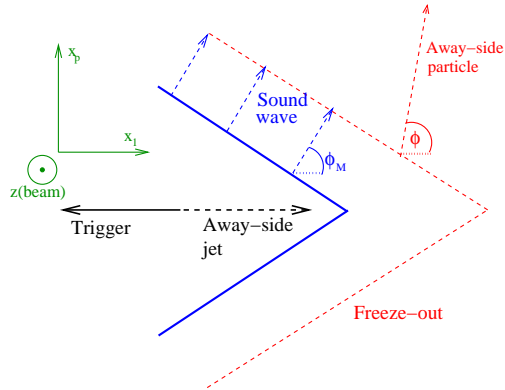,width=2.6in,clip=}
\caption{\label{coordinates}
(Color online) Schematic of the geometry used here.
The trigger jet corresponds to the heavy
quark produced near the surface in the $-\hat{x}$ direction
transverse to beam axis ``z''. The away side jet moves in $\hat{x}$ direction
with velocity $v$. The
comoving coordinate is $x_1=x-vt$ and the transverse radial coordinate
relative to this is $x_p$. A Mach wake (solid blue line) is
produced at azimuthal angle $\phi_M$ in the $x_1-x_p$ plane.
}
\label{convention}
\end{center}
\end{figure}

We now turn to the observable consequences of the AdS string drag stress model by assuming the Cooper-Frye hadronization scheme \cite{cooperfrye} with isochronous freeze-out discussed in section \ref{sechydro} \footnote{This is a strong model assumption
on top of the AdS calculus and will need much closer scrutiny
in the future. We have, however, experimented with different hadronization conditions and observables in \cite{us7} and found the results of this section, as well as the analogous results in section \ref{secfreezeqcd}, to be qualitatively the same.}. Our system of coordinates is explained in more detail in Fig.\ \ref{convention}.
As the discussion in section \ref{sechydro} has shown, the only way that a nontrivial angular correlation can arise in the soft degrees of freedom within the AdS/CFT string drag model is if we relax the linearized approximation, i.e., the formal $N_c,\,\lambda\to\infty$ but $\sqrt{\lambda}/N_c^2\to 0$ condition used to derive the stress and boldly extrapolate towards more ``physical'' parameters to  make contact with our QCD world.

\begin{figure}[b]\begin{center}
\epsfig{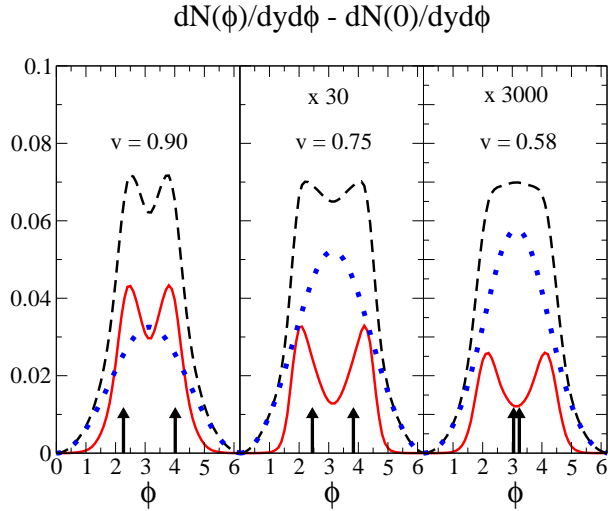}
\caption{\label{cfads}(Color online) Mid-rapidity azimuthal away-side associated angular
distribution from the Cooper-Frye freeze-out of the AdS/CFT string drag model $(T(x),\vec{U}(x))$ fields from \cite{gubser2}.
 Three cases for various heavy quark jet velocity
and associated hadron transverse momentum ranges,
$1: (v=0.9, p_T/\pi T_0=4$-$5$, $2: (v=0.75, p_T/\pi T_0=5$-$6$), and $3: (v=0.58, p_T/\pi T_0=6$-$7$), are compared. The short arrows show the
expected Mach angles. The red curves showing the double shoulder away side dip (conical) correlations are from the Neck region defined here
as where $\Delta\epsilon/\epsilon_{SYM}>0.3$ (see Fig.\ \ref{adscftmach}).
The blue curves result from integrating only in the far Mach zone outside
the Neck region and show no sign of the weak Mach wake seen in Fig.\ \ref{adscftmach} because the NO-GO freeze-out theorem (sec \ref{sechydro}) remains in force even for our
$N_c=3,\lambda=5.5$ downward extrapolation from the supergravity limit.
The sum total correlation exhibits a double shoulder correlation for $v<0.9$
arising from the chromo-viscous near zone that is however
unrelated to the Mach wakes seen in Fig. \ref{adscftmach}. }
\end{center}
\end{figure}

We computed $f(\phi;V)$ with $N_c=3, \lambda=5.5$ for $v=0.58,0.75,0.90$,
in a static uniform background
from the tables of $T^{00}$ and $T^{0i}$ (used in \cite{gubser2}). Our total CF volume is defined by $-14<X_1\,(\pi T_0)<1$, $0<X_{\perp}\,(\pi T_0)<14$, and $\varphi\in [0,2\pi]$. Here we define the head of the jet as the volume where $\xi>0.3$, which roughly corresponds to the region between $-1<X_1\,(\pi T_0)<1$ and $0<X_{\perp}\,(\pi T_0)<2$. We show results for the
azimuthal angular correlations in Fig. \ref{cfads}. The blue curves exclude the chromo-viscous Neck zone from the CF volume , the red Neck
curves only include the Neck zone approximated here by $\delta T^{00}(x) > 0.3\,\epsilon_{SYM}$.

As can be seen, only the red Neck curves display the double-peak structure while the ``Mach'' zone is too weak even in the $N_c=3$ extrapolation of AdS to produce a dip at $\phi=\pi$. For $v=0.9$ the two peaks from the Neck zone appear at angles accidentally similar to
the putative Mach cone angle.  Particles coming out of the ``Neck'' region will however probably not be thermalized (exponential, with the slope parameter given by temperature and
flow as in Eq. \ref{coopermach}), though it is difficult to say whether one could establish ``lack of thermalization'' conclusively from experimental data.

The fact that the ``cone-like signal'' arising from the neck region is not a {\em true} Mach cone has more direct phenomenological consequences: as shown in Fig.\ \ref{cfads}, the angle between the double peaks is relatively independent of the jet velocity $v$ (here tested for $v=0.58,0.75$ and $0.99$), which violates the Mach's law dependence (indicated by the small arrows). We propose that looking for deviations from Mach's law
for supersonic but not ultrarelativistic {\em identified heavy quark} jets could test this novel prediction of the AdS/CFT drag model.
Unlike light quark jets, even high momentum heavy quarks move at velocities significantly smaller than $c$.  A scan of Mach cone angles with the jet velocity should be, therefore, experimentally feasible. These results permit us to use the Mach-cone like signal to understand the ``phenomenology'' of AdS/CFT, since it is not at all obvious that {\em any} model will have a sizeable enough Neck region with the correct flow behavior. In fact, an explicit counter-example is the pQCD model proposed by \cite{neufeld}.

\section{Mach Cones in pQCD Coupled to Ideal Hydrodynamics \label{secpqcd}}

\subsection{Energy-Momentum Source Computed in pQCD\label{secsourceqcd}}

As before, we will decompose the energy-momentum tensor into a Head, Neck and Mach part according to Eq.\ (\ref{headneckmach}). However, in this section we focus on the case of a
pQCD plasma treated in the chromo-hydrodynamic limit \cite{neufeld}. The background stress tensor in this case is $T^{\mu\nu}_0={\rm diag}(\varepsilon_0,p_0,p_0,p_0)$, where
$\varepsilon_0=8\pi^2 T^4_0/15$ is the background energy density of gas of massless $SU(3)$ gluons (background temperature $T_0$) and $p_0=\varepsilon_0/3$. The role played by the
vacuum contribution to the energy-momentum tensor, $\delta T^{\mu\nu}_{Coul}(X)$, which is associated with the classical, non-Abelian Lorentz boosted Coulomb field created by the
fast moving parton, is to produce the anomalous response of the medium denoted by the Neck component. Here, we assume, as in section \ref{secfreezeads} and \cite{Dominguez:2008vd}, that the very near bare Coulomb field stress zone, in which $\delta T^{\mu\nu}_{Coul}\sim O(1/x^4)$,  is the self field stress of the heavy quark and does not fragment into associated hadrons.

The far zone ``Mach'' part of the stress can be expressed in terms of the axially symmetric local temperature $T(X)$ and fluid (Landau) flow velocity fields $U^\alpha(X)$ through the first-order Navier-Stokes stress form Eq.\ (\ref{hydrotensor}). Even though we will use the pQCD chromo-viscous source computed in linear response in \cite{neufeld,neufeldsource}, we will assume here the perfect fluid  to maximize the freeze-out azimuthal conical signature that is otherwise even more washed out when viscous dissipation is taken into account. Our aim here is not to fit RHIC data but rather to contrast weakly coupled and strongly coupled plasma response effects in the most idealized conditions of a uniform static plasma coupled to the external Lorentz contracted color $({\bf E}^a,{\bf B}^a)$ fields associated with a uniformly moving supersonic color charge.

Thus, while we assume in the background the perfect fluid $\eta/s=0$ limit of the full anomalous chromo-viscous equations derived in \cite{Asakawa:2006jn}, we also retain the anomalous diffusion stress Neufeld source \cite{neufeld,neufeldsource} that we rewrite in the more easily recognized Joule heating form (see Eqs.\ (6.2 - 6.11) of \cite{Asakawa:2006jn})
\begin{eqnarray}
\partial_\mu T^{\mu\nu}=\sigma^\nu= F^{\nu\alpha\,a}J_{\alpha}^a&=&
(F^{\nu\alpha\,a}C_{\alpha\beta\gamma}*F^{\beta\gamma\,a})
\end{eqnarray}
where $F^{\mu\nu\,a}(X)$ is the external Yang-Mills field tensor and
\begin{equation}
J^a(X)=\int d^4 K/(2\pi)^4  \exp(-iK\cdot X) J^a(K)
\end{equation}
 is the color current that is related via Ohm's law to $F^{\mu\nu\,a}(K)$ through the (diagonal in color) conductivity rank three tensor $C_{\mu\alpha\beta}(K)$. The $*$ denotes a convolution over the nonlocal non-static conductive dynamical response of the polarizable plasma.

The source is then \cite{neufeld}
\begin{equation}
\sigma^\nu=\int \frac{d^4{P} \,P^{\nu}}{(2\pi)^4}
\partial^\alpha_P \,D_{\alpha\beta}[X,P;F]\,\partial^\beta_P\;f(X,P)
\end{equation}
where $D_{\alpha\beta}$ is a quadratic form in the external field tensor $F^{\mu\nu\,a}$ components and $f(X,P)$ is the parton distribution function. In fact, the above and Eq.\ (28) of \cite{neufeld} factorize into the Joule heating $(F^{a}\cdot J^{a})^{\nu}$ form above.

The covariant generalization of Neufeld's source is easiest in Fourier decomposition with
$J_{\nu}^a(K)=C_{\nu\mu\alpha}(K) F^{\mu\alpha\,a}(K)$ and the color conductivity expression
derived in \cite{Selikhov:1993ns}
\begin{equation}
C_{\mu\alpha\beta}(K)=i g^2\int d^4 P\frac{P_\mu P_\alpha\,\partial^P_\beta}{
P\cdot K+i \,P\cdot U/\tau^*}f_0(P)
\end{equation}
where $f_0(P)=\left(N_c^2-1\right) \,G(P)$ is the effective
plasma equilibrium distribution with $G(P)=(2\pi^3)^{-1}\theta(P_0)\delta(P^2)/(e^{P_0/T}-1)$. Here, $U^\mu$ is the 4-velocity of the plasma as in Eq.\ (\ref{hydrotensor}). For an isotropic plasma $C_{\mu\alpha\beta}(K)=c_{\mu\alpha}(K)U_\beta$. In the long wavelength limit, $C^{\mu\alpha\beta}(K\rightarrow 0) =\tau^* m_D^2 \,g^{\mu\alpha}U^\beta /3$, where $m_D^2=g^2 T^2$ is the Debye screening mass for a noninteracting plasma of massless $SU(3)$ gluons in thermal equilibrium.

The relaxation  or decoherence time $\tau^*$ is of the generic form noted in \cite{Asakawa:2006jn}
\begin{equation}
\frac{1}{\tau^*}= \frac{1}{\tau_p} + \frac{1}{\tau_c} + \frac{1}{\tau_{an}}
\end{equation}
with $\tau_p\propto (\alpha_s^2 T\ln (1/\alpha_s))^{-1}$ being the collisional momentum relaxation time \cite{gyulvisc,Heinz:1985qe},
$\tau_c=(\alpha_s N_c T\ln (m_D/m_M))^{-1}$ being the color diffusion time defined in \cite{Selikhov:1993ns} in terms of the Debye electric and assumed $O(g^2T)$ magnetic screening masses, and $\tau_{an}\propto (m_D(\eta|\nabla U|/Ts)^{1/2})^{-1}$ being the anomalous strong electric and magnetic field relaxation time derived in Eq.\ (6.42) of \cite{Asakawa:2006jn}. Note that one can express
\begin{equation}
\tau_{an}=\frac{1}{gT} \frac{1}{\surd Kn(X)}
\label{anom}
\end{equation}
in terms of the local Knudsen number.
 However, because $\eta\propto
\tau^* sT$, Eq.(\ref{anom}) is really an implicit equation for $\tau_{an}$. Combining these relations and taking into account the uncertainty principle constraint discussed in 
section \ref{secads} and ($\tau^*\stackrel{>}{\sim} 1/\left(3T\right)$) \cite{gyulvisc}, 
\begin{equation}
 \frac{1}{\tau^*} = T \left(a_1\; g^4 \ln g^{-1}
+a_2\; g^2\ln g^{-1}  +  a_{3}\; g \surd
Kn \right) \stackrel{<}{\sim} 3T.
\end{equation}
As $Kn$ gets large, $\tau_{an}$ can get small even in the weak coupling limit. Thus, large gradients, which are the hallmark of the Neck zone, increase the importance of anomalous relaxation over color diffusion and collisional relaxation. Short relaxation times arise not only in the strong coupling but also in the weak coupling but classical field limits.


The full source $\sigma^\nu(X)$ was computed analytically by Neufeld in Ref.\ \cite{neufeldsource} and it reads
\begin{equation}
\sigma^{\mu}(X)=(\sigma^0,\bold{v}\sigma^0-\bold{\sigma}^v)\, .
\end{equation}
The components of the source term are reproduced below for convenience
\begin{equation}\label{one}
\sigma^0(t,{\bf x}) = d(t,{\bf x})\gamma \,v^2 \left[1 - \frac{x_1}{(x_1^2 + x_{\perp}^2)}\left(x_1 + \frac{\gamma\, v x_{\perp}^2}{\sqrt{x_{\perp}^2 + x_1^2 \gamma^2}}\right)\right]
\end{equation}
\begin{eqnarray}\label{two}
{\bf \sigma}_{v}(t,{\bf x}) &=& \left(\bold{x} - \bold{v} t\right)\frac{\alpha_s C\, m_D^2}{8\pi \left(x_1^2+x_{\perp}^2\right)^2}\nonumber\\
&&\hspace*{-1.5cm}\left[\frac{v^4 x_{\perp}^4+\left(x_1^2 \gamma ^2+x_{\perp} ^2\right) \left(2 x_1^2+\left(v^2+2\right)x_{\perp}^2/\gamma^2\right)}
{\left(x_1^2 \gamma^2+x_{\perp} ^2\right)^2}-\frac{2 v x_1}{\gamma  \sqrt{x_1^2 \gamma^2+x_{\perp} ^2}}\right]
\end{eqnarray}
where $\gamma=1/\sqrt{1-v^2}$, $\alpha_s=g^2/\left(4\pi\right)=1/\pi$, $m_D^2=g T$ and $C=4/3$ for a quark. These equations were obtained in the limit where the dielectric functions that describe the medium's response to the color fields created by the heavy quark were set to unity. The effects from medium screening on $\sigma^{\mu}$ were studied in detail in Ref.\ \cite{neufeldsource}. In our numerical calculations we used $x_{\perp\,max}=1/m_D$ as an infrared cutoff while the minimum lattice spacing naturally provided an ultraviolet cutoff. The background temperature was set to $T_0=0.2$ GeV.

This result holds for the weak coupling, long relaxation time $1/\tau^* =\epsilon\rightarrow 0^+$ limit. In this limit the color conductivity is large and Joule heating efficiently converts field energy and momentum into plasma heating and collective flow. The most interesting aspect of this for the present applications is the effect of highly inhomogeneous anisotropic Lorentz contracted Coulomb field of a uniformly moving heavy quark and color Casimir $4/3$ on the collective flow pattern imprinted on the plasma.

\subsection{Numerical Results for the Freeze-out \label{secfreezeqcd}}
\begin{figure}[h]\begin{center}
\epsfig{width=8cm,figure=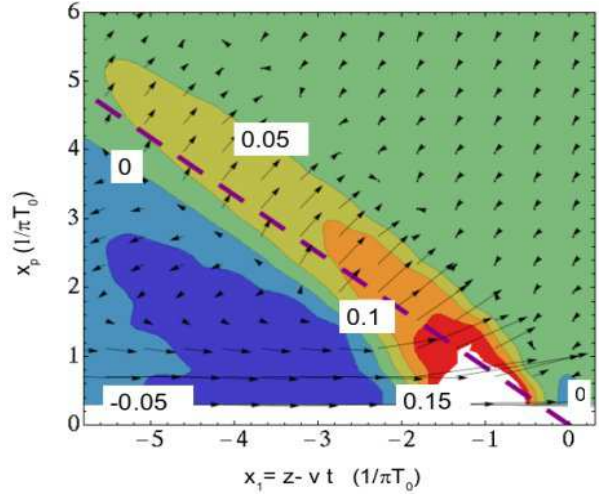}
\caption{\label{pqcdmach}(Color online) The relative local temperature perturbation
$\Delta T/T_0 = T(x_1, x_{\perp})/T_0 - 1$ and flow velocity profile
due to a heavy supersonic quark jet moving with $v = 0.9$
(speed of sound $c_s = 1/\sqrt{3}$). The results were obtained using
perfect fluid (3+1)D hydrodynamics in the presence of the
pQCD source term computed by Neufeld in \cite{neufeldsource}. The panel
shows the Mach wake (see purple dashed line) and trailing
shear column in the far zone as well as the Neck region (red)
near the jet. The heavy quark is at the origin of the coordinates.
The arrows show the direction and magnitude of the
flow. The numbers in the plot label the contours of constant
$\Delta T/T_0$. Note that non-Mach flow induced by Joule heating
is generated near the jet.}
\end{center}
\end{figure}

\begin{figure}[h]
\begin{center}
\epsfig{width=6.0cm,figure=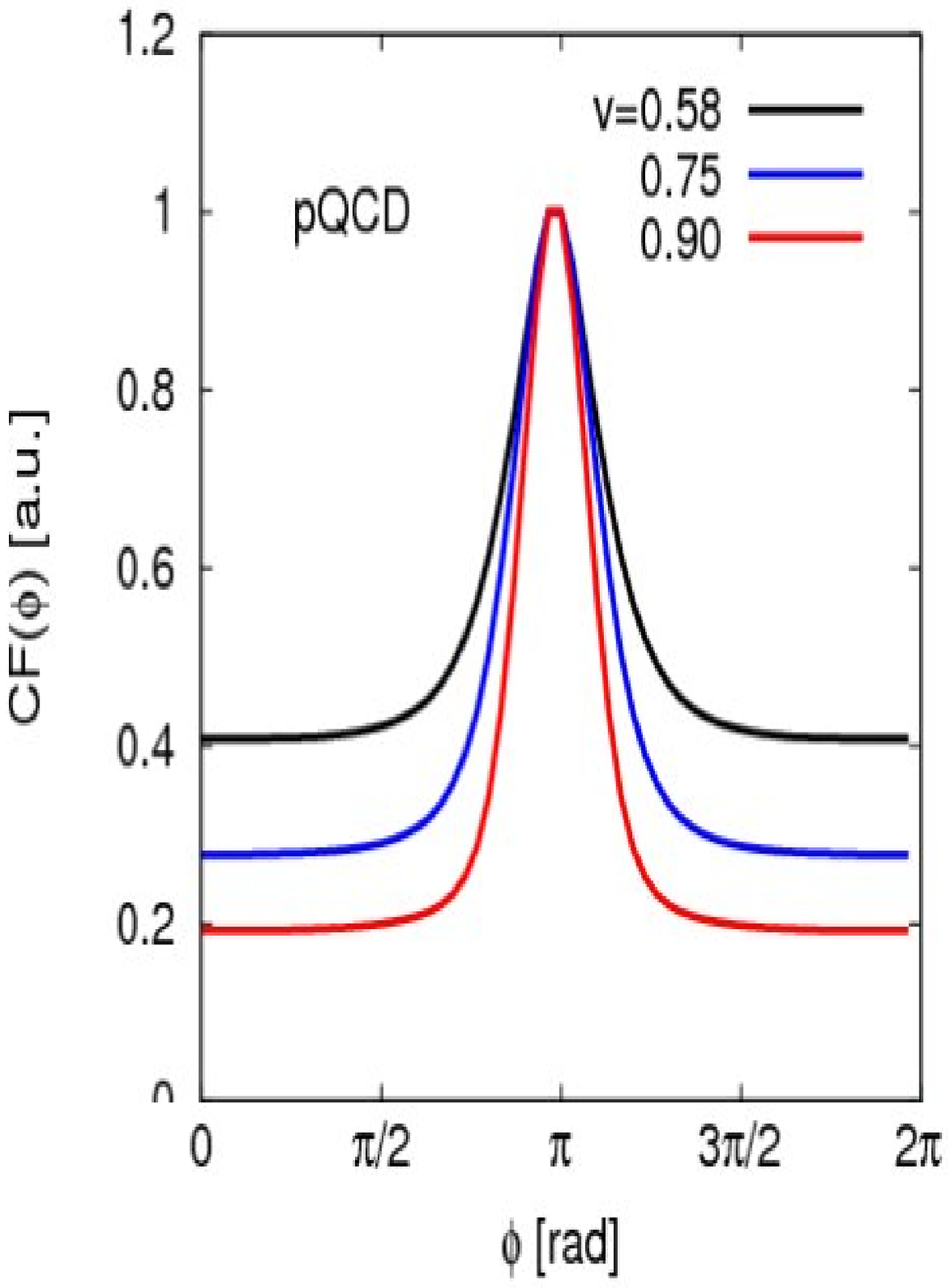}
\caption{\label{CFplot}
(Color online) Normalized and background subtracted azimuthal away side jet associated correlation after Cooper-Frye freeze-out (Eq. \ref{coopermach}) for pQCD. Here $f(\phi)=d^3N/dy dp_\perp d\phi$ is evaluated at $p_\perp=12.5 T_0$, and $y=0$. The black line is for $v=0.58$, the blue line for $v=0.75$, and for the red line $v=0.9$. Compare to the AdS/CFT result in Fig.\ \ref{cfads}.}
\end{center}
\end{figure}

We now turn to the observable consequences of the pQCD chromo-fluid flow.
The plot in Fig.\ \ref{pqcdmach} was obtained using the source of section \ref{secsourceqcd} integrated into the hydrodynamical code used in section \ref{sechydro}. The associated heavy quark jet is created in the beginning of the hydro evolution $t=0$ at $x_1 =-4.5$ fm and the freeze-out is done when the it reaches the origin of the coordinates, independently of the heavy quark's velocity (note that we do not include the trigger jet in our analysis). This provides a very rough description of the case in which a very energetic heavy quark punches through the medium.

Comparing Fig.\ \ref{pqcdmach} with Fig.\ \ref{adscftmach} it is apparent that the qualitative features displayed by both of these systems are very similar.  In particular, both
systems exhibit the Mach-like behavior when the distance from the source $\sim 5/T$.  The transverse flow in the Neck region, however, is considerably smaller in pQCD in
comparison to the AdS result found in section \ref{secfreezeads}.   We then expect, based on the study in section \ref{secfreezeads}, that the pQCD produced Mach cones are {\em not} observable.

We present normalized CF azimuthal distribution in the form of Eq. \ref{coopermach} in Fig.\ \ref{CFplot}. We computed $f(\phi)$  for $v=0.58,0.75,0.90$ in a static uniform background. The results for the azimuthal angular correlations for $p_T=8 \pi T_0=5$ GeV, and $y=0$ are shown in Fig.\ \ref{CFplot}. The pQCD angular distribution shows only a sharp peak at $\pi$ for all velocities while the AdS/CFT distribution displays the double-peak structure for all velocities shown in Fig. \ref{cfads}. Note, once again, that the peaks in the AdS/CFT correlation functions do not obey Mach's law since they do not come from the far-away linearized region.   The smaller transverse flow of the Neck produced by the pQCD source term ensures that the Mach contribution is washed out by the Cooper-Frye thermal smearing.

\section{Conclusions}

In conclusion, we have shown that Mach cones {\em in the fluid phase of the system} are a generic prediction of a medium characterized by low viscosity and a high opacity to hard probes. However, the signal seen in the experiment is likely to be considerably different from the naive expectation both because of the medium-induced thermal smearing at freeze-out and the presence of additional structures such as the diffusion wake.

We have illustrated these issues using examples taken from AdS/CFT and pQCD and shown that the critical requirement for a ``Mach-cone like signal'' is in fact the presence of a strong transverse flow in the {\em non-linear unthermalized} Neck region close to the source since this is most likely to survive freeze-out and yield a cone-like signal. However, the formation of strong transverse flow in the Neck region is far from assured. For instance, it is prominent in the AdS/CFT solution and missing in pQCD.

Our analysis is far from complete.  The most obvious physical aspects it misses are the underlying transverse flow (probably significant at timescales relevant to jet absorption), coalescence dynamics 
(certainly significant at momenta comparable to the away-side correlation 
used by experiment), and the dynamics of the freeze-out hypersurface. 
Efforts in this direction are ongoing \cite{us8} and more studies are needed before a conclusive link between the experimental result and the theory can be made.
Whille the Mach cone interpretations \cite{stoecker}  of
conical  correlations  are  intriguing and have focused attention
on moderate pT correlation observables,  the
observed conical correlations will only fulfill its promise
as a unique probe of the sQGP dynamics and its coupling to hard probes when
the many remaining open problems and caveats
discussed above are resolved. This effort is certainly warranted given the
potential
power of such correlation observables to resolve dynamical features
inaccessible with well studied and measured inclusive and bulk flow
observables.

\section*{Acknowledgments}

We thank S.~Gubser, S.~Pufu, and A.~Yarom for providing their numerical stress tables and I.~Mishustin, D.~Rischke and H.~St\"ocker for useful discussions. J.N. and M.G. acknowledge support from US-DOE Nuclear Science Grant No. DE-FG02-93ER40764. M.G. is grateful for DFG Mercator Gast
Professor support while on sabbatical at ITP/Goethe University. G.T. 
thanks the LOEWE and Alexander Von Humboldt foundations and Goethe
University for support. G.T. is extremely grateful to the organizers of the Krakow School of Theoretical Physics for providing him the opportunity and financial support to attend the school.



\end{document}